\documentclass[prd,nofootinbib,superscriptaddress]{revtex4-2}

\usepackage{amssymb}
\usepackage{amsmath}
\usepackage{xcolor}
\usepackage{tikz}
\usepackage{epsfig}
\usepackage{bbm}
\usepackage[colorlinks=true,linkcolor=blue,citecolor=blue,urlcolor=blue]{hyperref}

\allowdisplaybreaks

\definecolor{lime}{HTML}{A6CE39}
\definecolor{revisionorange}{HTML}{FF8C00}

\DeclareRobustCommand{\orcidicon}{
	\begin{tikzpicture}
	\draw[lime, fill=lime] (0,0) 
	circle [radius=0.2] 
	node[white] {{\fontfamily{qag}\selectfont \tiny ID}};
	\draw[white, fill=white] (-0.0625,0.095) 
	circle [radius=0.007];
	\end{tikzpicture}
	\hspace{-2mm}
}

\foreach \x in {A, ..., Z}{\expandafter\xdef\csname orcid\x\endcsname{\noexpand\href{https://orcid.org/\csname orcidauthor\x\endcsname}
			{\noexpand\orcidicon}}
}

\newcommand{\su}{\mathop{\mathfrak{su}}\nolimits}
\newcommand{\SU}{\mathop{\rm SU}\nolimits}
\newcommand{\1}{\mathbbm{1}}

\newcommand{\Tr}{\mathop{\rm Tr}\nolimits}

\newcommand{\mf}[1]{\textcolor{blue}{[Marco: #1]}}
\newcommand{\ag}[1]{\textcolor{red}{[Anish: #1]}}

\begin{document}

\title{Exact Lattice Identities and Continuum-Limit \\ Dyson--Schwinger Equations for Yang-Mills Theory}

\author{Arpan Chatterjee\orcidD{}}
\email{arpan.chatterjee@ut.ee}
\affiliation{F\"u\"usika Instituut, Tartu Ulikool,
  W.~Ostwaldi 1, EE-50411 Tartu, Estonia}

\author{Marco Frasca\orcidA{}}
\email{marcofrasca@mclink.it}
\affiliation{Rome, Italy}

\author{Anish Ghoshal\orcidB{}}
\email{a.ghoshal@sussex.ac.uk}
\affiliation{Department of Physics and Astronomy,
  University of Sussex, Brighton BN1 9RH, United Kingdom}

\author{Stefan Groote\orcidC{}}
\email{stefan.groote@ut.ee}
\affiliation{F\"u\"usika Instituut, Tartu Ulikool,
  W.~Ostwaldi 1, EE-50411 Tartu, Estonia}

\begin{abstract}
Starting from $\SU(N)$ on the lattice, we give a rigorous derivation of the
Dyson--Schwinger equations in the continuum limit. We formulate the
Dyson--Schwinger identities for the lattice Yang--Mills theory directly in
terms of the link variables $U_\mu(m)\in SU(N)$, exploiting the invariance of
the Haar measure under left group translations. This provides an exact lattice
derivation of the corresponding master equation for the Wilson action,
expressed through left-invariant Lie derivatives acting on individual links.
Because the construction is carried out directly on the compact gauge group,
it avoids the ambiguities associated with introducing Lie-algebra valued gauge
potentials as primary integration variables at finite lattice spacing. For
practical applications, in a second part we then break down the gauge degree
of freedom by choosing Feynman gauge. We analyze the continuum-limit form of
the resulting lattice identities and derive equations for the one- and
two-point connected functions. Under a further simplifying reduction, these
equations close to a tractable scalar system. Our results establish a direct
bridge between exact lattice identities and the functional equations commonly
used in continuum nonperturbative studies of Yang--Mills theory.
\end{abstract}

\maketitle

\newpage

\section{Introduction}
Nonperturbative Yang--Mills theory is controlled by identities that are exact,
but it is often studied through variables and approximations that obscure
where those identities come from. Lattice gauge theory and continuum
Dyson--Schwinger equations provide two complementary languages for this
problem: the first gives a gauge-invariant ultraviolet regulator built from
compact link variables, while the second gives a functional hierarchy for
Green functions and correlation functions \cite{DSEReview1,DSEReview2,Frasca:2015yva,Frasca:2024pmv}. 
A central goal of our analysis is to make the passage between these two languages 
as direct as possible.

The point is simple but important. On the lattice the fundamental integration
variables are not Lie-algebra valued gauge potentials. They are group-valued
links $U_\mu(m)\in\SU(N)$ integrated with the normalized Haar
measure~\cite{Montvay1994,Briceno:2025hca,Rothe2012}. In contrast, continuum Dyson--Schwinger
equations are often motivated by the formal translational invariance of a flat
gauge-field measure. At finite lattice spacing this flat-measure intuition is
not intrinsic: the exponential map has a nontrivial Jacobian, is not globally
one-to-one, and gauge-variant Green functions require a gauge choice and the
associated Faddeev--Popov structure.

The main result of the present work is an exact finite-lattice derivation of
the Dyson--Schwinger master identity directly on the compact group manifold.
The derivation uses left invariance of the Haar measure and left-invariant Lie
derivatives acting on a single link. Applied to the Wilson action, this
identity becomes a link-local equation in which the variation of a link is
expressed through the neighboring staple variables. Thus the Wilson lattice
action itself produces the lattice Dyson--Schwinger hierarchy without first
replacing the compact variables by gauge potentials.

This perspective also clarifies the relation to earlier work. Schwinger--Dyson
equations and Ward--Takahashi identities on finite lattices, including the
role of gauge fixing, were analyzed using lattice derivatives in
Refs.~\cite{Kerler:1981sd,Kerler:1981prd}. Closely related
integration-by-parts ideas underlie Wilson-loop and loop-equation approaches,
starting with the Makeenko--Migdal equations and continuing in modern rigorous
treatments of Yang--Mills measures~\cite{Makeenko:1979,Driver:2019}. Our
contribution is to give an explicit link-by-link master equation in compact
$\SU(N)$ variables based on the Wilson action and to trace how its lower
sectors connect to the continuum functional equations used in nonperturbative
analyses.

A useful feature of this formulation is that it cleanly separates what is
exact from what is additional. The Haar-measure identity and the Wilson-action
master equation are exact at finite lattice spacing. The subsequent continuum
expansion is a formal comparison with gauge-fixed gluonic Dyson--Schwinger
equations, and the scalar closure considered later is an additional reduction
ansatz for a special color-Lorentz sector. Keeping these layers separate is
essential for using the construction responsibly, and it is also what makes
the framework useful: one can improve the continuum reduction, gauge fixing,
or truncation scheme without changing the exact lattice identity from which
the analysis starts.

The broader motivation is to build truncation strategies that inherit more of
the lattice theory's compact-group structure. Such strategies could provide a
more controlled bridge between lattice regularization, functional
Green-function methods, Wilson-loop identities, and nonperturbative continuum
physics. In particular, the link-local master equation derived here gives a
starting point for future gauge-invariant formulations, ghost-complete
gauge-fixed reductions, and lattice-informed continuum truncations.

The paper is organized as follows. Section~II formulates the Haar-measure
integration-by-parts identity for compact link variables and fixes the
group-theory conventions. Section~III applies this identity to the Wilson
action and derives the exact lattice master equation. Section~IV analyzes the
formal continuum expansion and the gauge-fixed interpretation of the one- and
two-point equations. Section~V presents the scalar reduction ansatz and the
resulting closed lower-sector equations. Appendix~A records the explicit
plaquette derivative used in the continuum expansion.

\section{Haar-Measure Integration by Parts on the Lattice}
A common way to connect lattice variables with continuum Dyson--Schwinger (DS)
equations is to introduce gauge-field variables $A_\mu^a(m)$ through the local
parametrization $U_\mu(m)=\exp(iga A_\mu^a(m)T^a)$ and then to reason in terms
of a formally flat measure $\mathcal{D}[A]$. This may be useful in
perturbative or continuum-oriented settings, but it hides several
finite-spacing issues that are absent in a formulation based directly on the
compact group variables:
\begin{enumerate}
\item The change of variables $A_\mu^a(m)\mapsto U_\mu(m)$ introduces a
non-trivial Jacobian $J[A]$ that depends on the fields and cannot be neglected
at finite lattice spacing $a$.
\item The exponential map $\exp:\su(N)\to\SU(N)$ is not globally injective;
configurations with $\|gaA\|\sim\pi$ contribute significantly to the path
integral, particularly in the confined phase.
\item The flat measure $\mathcal{D}[A]$ requires gauge fixing
(Faddeev--Popov procedure, ghosts) to be well defined, while the Haar measure
$\mathcal{D}[U]$ is gauge invariant and well defined on the compact group
without any of this.
\end{enumerate}
For the finite lattice theory, the intrinsic formulation is therefore to
derive the DS identities directly in terms of the link variables
$U_\mu(m)\in\SU(N)$, using the invariance of the Haar measure under left group
translations. This gives exact lattice identities before any continuum
parametrization is introduced. In particular, no Jacobian from a change of
variables appears at this stage, and no gauge fixing is required merely to
define the compact-group path integral.

For a link $U_\mu(m)\in\SU(N)$, the integration measure is the normalized
Haar measure $dU_\mu(m)$ on $\SU(N)$, satisfying
\begin{eqnarray}
\int_{\SU(N)}dU&=&1,\label{eq:haar_norm}\\
\int_{\SU(N)}f(VU)dU&=&\int_{\SU(N)}f(U)dU\qquad\forall\,V\in\SU(N),
  \label{eq:haar_left}\\
\int_{\SU(N)}f(UV)dU&=&\int_{\SU(N)}f(U)dU\qquad\forall\,V\in\SU(N).
  \label{eq:haar_right}
\end{eqnarray}
The full lattice measure is
\begin{equation}\label{eq:full_measure}
  {\cal D}[U]=\prod_{m,\mu}dU_\mu(m).
\end{equation}
This compact-group measure is exact at finite lattice spacing: no Jacobian,
ghost determinant, or gauge-fixing condition is required for the measure
itself to be well defined. For $U\in\SU(N)$ and $T^a\in\su(N)$ with
$[T^a,T^b]=if_{abc}T^c$, the (left) Lie derivative is given by
\begin{equation}
{\cal L}^a F(U):=\frac{d}{d\epsilon}F\big(e^{i\epsilon T^a}U\big)
  \Big|_{\epsilon=0}.
\end{equation}
Throughout the paper we use Hermitian generators of $\su(N)$ in the
fundamental representation, normalized as $\Tr(T^aT^b)=\delta^{ab}/2$, with
$[T^a,T^b]=if^{abc}T^c$ and $\{T^a,T^b\}=\delta^{ab}\1/N+d^{abc}T^c$. With
this convention, all factors of $i$ in the left Lie derivative follow from
$\delta U=i\epsilon T^aU$. We also reserve $m,n,o$ for lattice sites and
$\mu,\nu,\rho$ for directions; repeated color and Lorentz indices are summed
unless explicitly stated otherwise. The key identity underlying all DS
equations follows from left-invariance of the Haar measure, formulated in
Eq.~(\ref{eq:haar_left}). As a consequence, for any differentiable functional
$F[U]$ on the lattice one has
\begin{equation}\label{eq:IBP}
  \int{\cal D}[U]\,{\cal L}^a_\mu(m)F[U]=0,
\end{equation}
where ${\cal L}^a_\mu(m)$ denotes the left Lie derivative acting on the link
$U_\mu(m)$ while leaving all other links unchanged,
\begin{equation}
  {\cal L}^a_\mu(m)F[U]:=\frac{d}{d\epsilon}F(\ldots,e^{i\epsilon T^a}U_\mu(m),
  \ldots)\Big|_{\epsilon=0}.
\end{equation}
This is the finite-lattice counterpart of the formal continuum identity
$\int {\cal D}A\,\delta F/\delta A^a_\mu(x)=0$. The lattice statement is
stronger in the present context because it is an ordinary integration-by-parts
identity on a compact finite-dimensional group manifold and therefore holds
without a continuum or small-field approximation. No gauge fixing is required
for the normalized Haar measure or for the integration-by-parts
identity~(\ref{eq:IBP}). This statement concerns the definition of the
finite-lattice functional integral and is independent of the choice of
insertion $F[U]$. Its physical interpretation depends, however, on the
insertion. Gauge-invariant insertions, including Wilson loops and products of
plaquette traces, define observables directly in the unfixed theory.
Gauge-variant insertions are mathematically admissible in the identity, but
their unfixed expectation values are averaged over the local gauge orbit and
cannot in general be identified with gluon Green functions.


\section{Wilson Action and Exact Lattice Master Equation}
For Wilson lattice Yang--Mills theory, and up to an additive constant
independent of the links, we use the following convention for the Wilson
action:
\begin{eqnarray}
S_W[U]&=&-\frac\beta{N}\sum_m\sum_{\mu,\nu\ne\mu}U_{\mu\nu}(m)
  \ =\ -\frac\beta{2N}\sum_m\sum_{\mu,\nu<\mu}(U_{\mu\nu}(m)+U_{\nu\mu}(m))
  \ =\nonumber\\
  &=&-\frac\beta{4N}\sum_m
  \left(\sum_{\mu,\nu<\mu}(U_{\mu\nu}(m)+U_{\nu\mu}(m))
  +\sum_{\nu,\mu>\nu}(U_{\mu\nu}(m)+U_{\nu\mu}(m))\right)\ =\nonumber\\
  &=&-\frac\beta{4N}\sum_m
  \left(\sum_{\mu,\nu<\mu}(U_{\mu\nu}(m)+U_{\nu\mu}(m))
  +\sum_{\mu,\nu>\mu}(U_{\nu\mu}(m)+U_{\mu\nu}(m))\right)\ =\nonumber\\
  &=&-\frac\beta{4N}\sum_m\sum_{\mu,\nu\ne\mu}(U_{\mu\nu}(m)+U_{\nu\mu}(m))
  \ =\ -\frac\beta{4N}\sum_m\sum_{\mu,\nu\ne\mu}(U_{\mu\nu}(m)
  +U_{\mu\nu}^\dagger(m))
\end{eqnarray}
with $\beta=2N/g^2$, where the plaquette in the $(\mu,\nu)$ plane is given as
\begin{equation}
U_{\mu\nu}(m)=\Tr\left[U_\mu(m)U_\nu(m+\hat\mu)U_\mu^\dagger(m+\hat\nu)
  U_\nu^\dagger(m)\right]=U_{\nu\mu}^\dagger(m).
\end{equation}
With this orientation convention, one has
$U_{\nu\mu}(m)=U_{\mu\nu}^\dagger(m)$, and the trace of each plaquette is
gauge invariant. The staple notation below isolates all plaquettes that
contain a fixed link $U_\mu(m)$, which is the only link varied by
${\cal L}_{\mu,m}^a$.\par\noindent
\parbox[b]{9truecm}{\hsize=9truecm\noindent
Considering two neighbouring points $m$ and $m+\hat\mu$ in $\mu$-direction,
the plaquette can be decomposed into a (conjugate) link $U_\mu^\dagger(m)$ and
the forward staple
\begin{equation}
\Sigma_{\mu|\nu}^\uparrow(m):=U_\nu(m)U_\mu(m+\hat\nu)U_\nu^\dagger(m+\hat\mu),
\end{equation}
with
$U_{\mu\nu}^\dagger(m)=\Tr[\Sigma_{\mu|\nu}^\uparrow(m)U_\mu^\dagger(m)]$. The
link can also be extracted from the opposite edge of the plaquette, leaving
the backward staple
\begin{equation}
\Sigma_{\mu|\nu}^\downarrow(m):=U_\nu^\dagger(m-\hat\nu)U_\mu(m-\hat\nu)
  U_\nu(m+\hat\mu-\hat\nu)
\end{equation}}\hfill
\parbox[b]{7truecm}{\hsize=7truecm\noindent
\hbox to7truecm{\hss\epsfig{figure=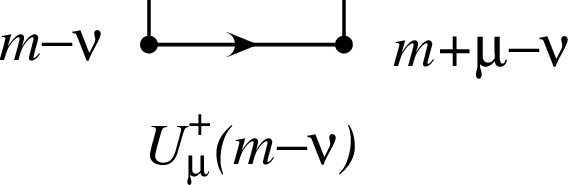, scale=0.5}\hss}}
with
$U_{\mu\nu}(m-\hat\nu)=\Tr[\Sigma_{\mu|\nu}^\downarrow(m)U_\mu^\dagger(m)]$.
For a fixed value of $\mu$, we obtain
\begin{eqnarray}
\lefteqn{\sum_m\sum_{\nu\ne\mu}
  \left(U_{\mu\nu}(m)+U_{\mu\nu}^\dagger(m)\right)\ =}\nonumber\\
  &=&\frac12\sum_m\sum_{\nu\ne\mu}\left(U_{\mu\nu}(m)+U_{\mu\nu}^\dagger(m)
  +U_{\mu\nu}(m-\hat\nu)+U_{\mu\nu}^\dagger(m-\hat\nu)\right)\ =\nonumber\\
  &=&\frac12\sum_m\sum_{\nu\ne\mu}\Tr\left[
  U_\mu(m)\Sigma_{\mu|\nu}^{\uparrow\dagger}(m)+\Sigma_{\mu|\nu}^\uparrow(m)
  U_\mu^\dagger(m)+\Sigma_{\mu|\nu}^\downarrow(m)U_\mu^\dagger(m)
  +U_\mu(m)\Sigma_{\mu|\nu}^{\downarrow\dagger}(m)\right]\ =\nonumber\\
  &=&\frac12\sum_m\Tr\left[U_\mu(m)\Sigma_\mu^\dagger(m)+\Sigma_\mu(m)
  U_\mu^\dagger(m)\right],
\end{eqnarray}
where forward and backward staple are then combined into the staple
\begin{equation}
\Sigma_\mu(m):=\sum_{\nu\ne\mu}
  \left(\Sigma_{\mu|\nu}^\uparrow(m)+\Sigma_{\mu|\nu}^\downarrow(m)\right).
\end{equation}
Based on the fundamental integration-by-parts identity~(\ref{eq:IBP}), we
calculate the Lie derivative of the Wilson action that is given by
\begin{equation}\label{leftLie}
{\cal L}_{\mu,m}^aS_W[U]=-\frac{i\beta}{8N}\Tr\left[T^aU_\mu(m)
  \Sigma_\mu^\dagger(m)-\Sigma_\mu(m)T^aU_\mu^\dagger(m)\right].
\end{equation}
Using the generating functional
\begin{equation}
{\cal Z}[J]=\int{\cal D}[U]\exp\left(-S_W[U]+\sum_{n,\nu,b}J^\nu_b(n)
  \Tr[T^bU_\nu(n)]\right)
\end{equation}
with summation over $\mu$ and $a$ implied, we can generate connected
correlation functions by calculating the variations of the effective action
${\cal W}[J]=\ln{\cal Z}[J]$ with respect to $J$,
\begin{eqnarray}\label{Gdef}
G_\mu^a(m)&:=&\frac{\delta{\cal W}}{\delta J^\mu_a(m)}
  \ =\ \langle\Tr[T^aU_\mu(m)]\rangle\ =:\ \langle\Tr[T^aU_\mu(m)]\rangle_c,
  \nonumber\\
G_{\mu\nu}^{ab}(m,n)&:=&\frac{\delta^2{\cal W}}{\delta
  J^\mu_a(m)\delta J^\nu_b(n)}=\langle\Tr[T^aU_\mu(m)]\Tr[T^bU_\nu(n)]\rangle
  +\strut\nonumber\\&&\strut
  -\langle\Tr[T^aU_\mu(m)]\rangle\langle\Tr[T^bU_\nu(n)]\rangle
  \ =:\ \langle\Tr[T^aU_\mu(m)]\Tr[T^bU_\nu(n)]\rangle_c,\qquad
\end{eqnarray}
where the label $c$ indicates the connected correlation. The quantities
generated by the source in Eq.~(\ref{Gdef}) are formal link-source cumulants.
Because $\Tr[T^aU_\mu(m)]$ is not gauge invariant, these cumulants are not, by
themselves, physical observables of the unfixed lattice theory. Their
interpretation as gluon Green functions requires a choice of gauge and a
corresponding gauge-fixed generating functional. The exact Haar-measure
identity remains valid independently of this later choice. Because the source
couples to $\Tr[T^aU_\mu]$, these objects are gauge variant. They should be
interpreted either as formal link-source derivatives of the exact lattice
functional or, after a gauge choice, as precursors of gauge-fixed gluon Green
functions. The goal is to derive lattice difference equations for these
connected quantities and then compare their formal continuum limit with the
corresponding differential equations for low-order Green functions. As usual,
these equations are derived from the tower of Dyson--Schwinger equations, and
the master equation is derived by acting the Lie derivative to the generating
functional and using the Leibniz rule,
\begin{equation}
\langle{\cal L}_{\mu,m}^a{\cal S}_W[U]\rangle_J=\left\langle{\cal L}_{\mu,m}^a
  \left[\sum_nJ_\nu^b(n)\Tr[T^bU_\nu(n)]\right]\right\rangle_J
  =:\left\langle\sum_{\nu\ne\mu}{\cal L}^a_\mu U_{\mu\nu}(m)\right\rangle,
\end{equation}
so explicitly (and with no summation over $\mu$ implied)
\begin{eqnarray}\label{master}
\lefteqn{-\frac{i\beta}{8N}\left\langle\Tr\left[T^aU_\mu(m)
  \Sigma_\mu^\dagger(m)-\Sigma_\mu(m)T^aU_\mu^\dagger(m)\right]
  \right\rangle_J\ =}\nonumber\\
  &=&J_{\mu,m}^b\left\langle\frac1{2N}\delta^{ab}\Tr[U_\mu(m)]
  +\frac12(d^{abc}-if^{abc})\Tr[T^cU_\mu(m)]\right\rangle_J,
\end{eqnarray}
where for the last step we used ${\cal L}_{\mu,m}^a\Tr[T^bU_\nu(n)]
=i\delta_{mn}\eta_{\mu\nu}\Tr[T^bT^aU_\nu(n)]$ and
\begin{equation}
T^bT^a=\frac1{2N}\delta^{ab}\1+\frac12(d^{abc}-if^{abc})T^c.
\end{equation}
The Lie derivative ${\cal L}^a_\mu U_{\mu\nu}(m)$ of the plaquette for fixed
directions $\hat\mu$ and $\hat\nu$ is calculated in Appendix~A. As a first
algebraic check, one may take the localization limit
$\xi_\mu^a(m+\ldots)\to\xi_\mu^a(m)=:\xi_\mu^a$, in which all shifted fields
are identified at the same point. This gives
\begin{eqnarray}
{\cal L}^a_\mu U_{\mu\nu}(m)
  &=&-\frac{a^3g^3}6\left\{12\xi^3(m)\Tr(T^a\eta_\nu^2\eta_\mu)
  -24\xi^3(m)\Tr(T^a\eta_\nu\eta_\mu\eta_\nu)
  +12\xi^3(m)\Tr(T^a\eta_\mu\eta_\nu^2)\right\}\ =\nonumber\\
  &=&-2a^3g^3\xi^3(m)\left\{\Tr(T^a\eta_\nu^2\eta_\mu)
  -2\Tr(T^a\eta_\nu\eta_\mu\eta_\nu)+\Tr(T^a\eta_\mu\eta_\nu^2)\right\}
  \ =\nonumber\\
  &=&-2a^3g^3\xi^3(m)\Tr(T^a[[\eta_\mu],\eta_\nu],\eta_\nu])
  \ =\ a^3g^3\xi^3(m)f_{bcd}f_{dea}\eta_\mu^b\eta_\nu^c\eta_\nu^e.
\end{eqnarray}

\section{Continuum Expansion and Equations for One- and Two-Point Functions}
We now change the level of description. The results of Secs.~II and III are
exact identities of the unfixed compact-link theory. Instead, the comparison
in this section is gauge dependent. A complete covariant-gauge continuum
generating functional contains, in addition to the gauge-fixing term, the
Faddeev-Popov determinant or, equivalently, ghost fields. The associated gluon
Dyson-Schwinger equation is coupled to ghost correlation functions and is
constrained by Slavnov--Taylor identities. The purpose of the present section
is to extract the purely gluonic projection of the formal continuum expansion
by adding the Feynman-gauge quadratic term and omitting the explicit ghost
sector. The equations below should therefore not be interpreted as the
complete gauge-fixed Yang--Mills Dyson--Schwinger hierarchy. They are
conditional, ghost-omitted comparison equations used to illustrate how the
exact lattice master identity reproduces the tensor structure of the gluonic
lower sectors.

The source used in Eq.~(\ref{Gdef}) generates gauge-variant link correlators.
Consequently, the exact finite-lattice Haar identity derived above does not by
itself define gauge-field Green functions in a fixed gauge. In the
continuum-oriented comparison below we therefore supplement the formal
expansion by a Feynman-gauge condition. A complete gauge-fixed formulation
would include the corresponding Faddeev--Popov determinant and ghost sector,
as in the standard continuum treatment~\cite{Faddeev:1967,Slavnov:1972fg,%
Taylor:1971ff}. In our analysis we use the gauge-fixing term only to display
the formal continuum form of the gluonic part of the lower Dyson--Schwinger
equations; the ghost-sector completion is left outside the present reduction.
For consistency with the scalar-field analysis of Ref.~\cite{Frasca:2026dyg},
we are interested in the continuum limit $a\to 0$. The Taylor series expansion
into the lattice spacing $a$ results in
\begin{eqnarray}
\lefteqn{{\cal L}^a_\mu U_{\mu\nu}/a^3\ =\ \sum_{\nu\ne\mu}\Bigg\{
  g\left[\partial^\nu\partial_\nu\xi_\mu^a
  -\partial^\nu\partial_\mu\xi_\nu^a\right]+\strut}\nonumber\\&&\strut
  -g^2f_{abc}\left[(\partial_\mu\xi^\nu_b)\xi_\nu^c
  -2(\partial^\nu\xi_\mu^b)\xi_\nu^c+(\partial^\nu\xi_\nu^b)\xi_\mu^c\right]
  +g^3f_{abe}f_{cde}\xi^\nu_b\xi_\nu^c\xi_\mu^d+O(g^4)\Bigg\}.
\end{eqnarray}
Together with the gauge fixing term in Feynman gauge, given by
\begin{equation}
2g\sum_\nu\partial^\nu\partial_\mu\xi_\nu^b\Tr[T^aT^b]
  =g\sum_\nu\partial^\nu\partial_\mu\xi_\nu^a
\end{equation}
(note that the sum is running here over all $\nu$), the second term in the
first row of ${\cal L}^a_\mu U_{\mu\nu}(m)/a^3$ is cancelled, and the term
$\nu=\mu$ is added. Therefore, we end up with
\begin{eqnarray}\label{tracedU}
\lefteqn{{\cal L}^a_\mu U_{\mu\nu}(m)/a^3\ =\ g\partial^2\xi_\mu^a(m)
  +\strut}\nonumber\\&&\strut
  -g^2f_{abc}\Big[(\partial_\mu\xi^\nu_b(m))\xi_\nu^c(m)
  -2(\partial^\nu\xi_\mu^b(m))\xi_\nu^c(m)
  +(\partial^\nu\xi_\nu^b(m))\xi_\mu^c(m)\Big]+\strut\nonumber\\&&\strut
  +g^3f_{bcd}f_{dea}\xi_\mu^b(m)\xi_\nu^c(m)\xi^\nu_e(m)+O(g^4).
\end{eqnarray}
Before inserting the result obtained so far into the Dyson--Schwinger master
equation~(\ref{master}), note that $2\Tr[T^aU_\mu(m)]=\xi_\mu^a(m)$, and the
system of equations~(\ref{Gdef}) for the $n$-point functions can be inverted
to
\begin{eqnarray}\label{Green}
\lefteqn{\langle i\xi_\mu^a(m)\rangle\ =\ G_\mu^a(m),\quad
\langle i\xi_\mu^a(m)i\xi_\nu^b(n)\rangle\ =\ G_{\mu\nu}^{ab}(m,n)
  +G_\mu^a(m)G_\nu^b(n),}\nonumber\\
\lefteqn{\langle i\xi_\mu^a(m)i\xi_\nu^b(n)i\xi_\rho^c(o)\rangle
  \ =\ G_{\mu\nu\rho}^{abc}(m,n,o)+G_\mu^a(m)G_{\nu\rho}^{bc}(n,o)
  +\strut}\nonumber\\&&\strut\qquad\qquad\qquad
  +G_\nu^b(n)G_{\mu\rho}^{ac}(m,o)
  +G_\rho^c(o)G_{\mu\nu}^{ab}(m,n)
  +G_\mu^a(m)G_\nu^b(n)G_\rho^c(o)\qquad
\end{eqnarray}
Therefore, inserting into Eq.~(\ref{master}) results in
\begin{eqnarray}
\lefteqn{i\partial^2G_\mu^a(m)\ =\ igf_{abc}\Big[
  \Big(\partial_\mu^mG_{b\nu}^{\nu c}(m,n)
  -2\partial^\nu_mG_{\mu\nu}^{bc}(m,n)+\partial^\nu_mG_{\nu\mu}^{bc}(m,n)
  \Big)\Big|_{n=m}+\strut}\nonumber\\&&\strut\quad
  +(\partial_\mu G^\nu_b(m))G_\nu^c(m)-2(\partial^\nu G_\mu^b(m))G_\nu^c(m)
  +(\partial^\nu G_\nu^b(m))G_\mu^c(m)\Big]+\strut\nonumber\\&&\strut
  +ig^2f_{bcd}f_{dea}\Big[G_{\mu\nu e}^{bc\nu}(m,m,m)+G_{\mu\nu}^{bc}(m,m)
  G_\nu^e(m)+G_{\mu e}^{b\nu}(m,m)G_\nu^c(m)+\strut\nonumber\\&&\strut\qquad
  +G_{\nu e}^{c\nu}(m,m)G_\mu^b(m)+G_\mu^b(m)G_\nu^c(m)G^\nu_e(m)\Big].
\end{eqnarray}
In the complete gauge-fixed hierarchy, the right-hand side of the one-point
equation also contains a ghost-current contribution. With the conventional
Faddeev--Popov action
$S_{\rm gh}=\int d^Dx\,\bar c^a\partial^\mu D_\mu^{ab}c^b$, its schematic form
is
\begin{equation}\label{eq:ghost-current}
  {\cal G}_\mu^a(x)=gf^{abc}\left\langle(\partial_\mu\bar c^b(x))c^c(x)
  \right\rangle_J,
\end{equation}
up to the overall sign and placement of the derivative associated with the
convention chosen for $S_{\rm gh}$. In the present gluonic projection we set
${\cal G}_\mu^a=0$ as a truncation. This is an additional assumption and is
not a consequence of Haar-measure invariance or of the Feynman-gauge condition.

Here the superscript $m$ on a derivative indicates that the derivative acts on
the first argument of the two-point function. In the following expressions we
adopt this convention implicitly and suppress the label when no ambiguity can
arise. Taking a variational derivative with respect to the current
$J_\nu^b(n)$ generates the insertion $iag\xi_\nu^b(n)/2$ on the left-hand side
of the master equation~(\ref{master}), while the source derivative on the
right-hand side gives the contact term
$a^4\delta^{ab}\eta_{\mu\nu}\delta_{mn}/2$. After canceling the common factor
$a^4/2$ and multiplying by $8iN/\beta=4ig^2$, one obtains
\begin{eqnarray}
\lefteqn{4ig^2\delta^{ab}\eta_{\mu\nu}\delta_{mn}\ =\ -g^2\Big[
  \partial^2_mG_{\mu\nu}^{ab}(m,n)+(\partial^2G_\mu^a(m))G_\nu^b(n)
  \Big]+\strut}\nonumber\\&&\strut
  +g^3f_{acd}\Big[\partial_\mu G_{c\rho\nu}^{\rho db}(m,m,n)
  -2\partial^\rho G_{\mu\rho\nu}^{cdb}(m,m,n)
  +\partial^\rho G_{\rho\mu\nu}^{cdb}(m,m,n)+\strut\nonumber\\&&\strut\qquad
  +\Big(\partial_\mu G_{c\rho}^{\rho d}(m,m)
  -2\partial^\rho G_{\mu\rho}^{cd}(m,m)
  +\partial^\rho G_{\rho\mu}^{cd}(m,m)\Big)G_\nu^b(n)
  +\strut\nonumber\\&&\strut\qquad
  +\Big(\partial^\rho G_{c\nu}^{\rho b}(m,n)
  -2\partial^\rho G_{\mu\nu}^{cb}(m,n)
  +\partial^\rho G_{\rho\nu}^{cb}(m,n)\Big)G_\rho^d(m)
  +\strut\nonumber\\&&\strut\qquad
  +(\partial_\mu G^\rho_c(m))G_{\rho\nu}^{db}(m,n)
  -2(\partial^\rho G_\mu^c(m))G_{\rho\nu}^{db}(m,n)
  +(\partial^\rho G_\rho^c(m))G_{\mu\nu}^{db}(m,n)
  +\strut\nonumber\\&&\strut\qquad
  +\Big((\partial_\mu G^\rho_c(m))G_\rho^d(m)
  -2(\partial^\rho G_\mu^c(m))G_\rho^d(m)
  +(\partial^\rho G_\rho^c(m))G_\mu^d(m)\Big)G_\nu^b(n)\Big]
  +\strut\nonumber\\&&\strut
  -g^4f_{cde}f_{efa}\Big[G_{\mu\rho}^{cd}(m,m)G_{f\rho}^{\rho b}(m,n)
  +G_{\mu f}^{c\rho}(m,m)G_{\rho\nu}^{db}(m,n)
  +G_{\rho f}^{d\rho}(m,m)G_{\mu\nu}^{cb}(m,n)+\strut\nonumber\\&&\strut\quad
  +G_\mu^c(m)G_\rho^d(m)G_{f\nu}^{\rho b}(m,n)+G_\mu^c(m)G^\rho_f(m)
  G_{\rho\nu}^{db}(m,n)+G_\rho^d(m)G^\rho_f(m)G_{\mu\nu}^{cb}(m,n)
  +\strut\nonumber\\&&\strut\quad
  +\left(G_\mu^c(m)G_{\rho f}^{d\rho}(m,m)+G_\rho^d(m)G_{\mu f}^{c\rho}(m,m)
  +G^\rho_f(m)G_{\mu\rho}^{cd}(m,m)\right)G_\nu^b(n)
  +\strut\nonumber\\&&\strut\quad
  +G_\mu^c(m)G_\rho^d(m)G^\rho_f(m)G_\nu^b(n)\Big].
\end{eqnarray}
In the complete two-point equation, differentiating ${\cal G}_\mu^a$ with
respect to the gluon source also generates mixed ghost-gluon correlation
functions. These terms are omitted together with the ghost current in the
present projection.

\section{Scalar Reduction Ansatz}
The scalar system derived below is a reduction of the ghost-omitted gluonic
projection of Sec.~IV. It is not a closure of the complete gauge-fixed
Yang--Mills hierarchy. A ghost-complete scalar reduction would contain the
projected ghost current and mixed ghost-gluon functions and would have to
satisfy the corresponding Slavnov-Taylor identities.

The full hierarchy displayed above is not closed: the equation for an
$n$-point function couples to higher connected functions. To obtain an
analytically tractable lower-sector system, we now impose a scalar reduction
ansatz
\[
G_\mu^a(m)=\eta_\mu^a\phi(m),\qquad
G_{\mu\nu}^{ab}(m,n)=\eta_{\mu\nu}\delta^{ab}G(m,n).
\]
The constants $\eta_\mu^a$ specify a chosen color-Lorentz projection and are
normalized by Eq.~(\ref{ortho}). For the closure of the reduced equations we
use the same coincident-point Gaussianity criterion employed in the discrete
scalar-field analysis of Ref.~\cite{Frasca:2026dyg}. In the present
Yang--Mills setting this criterion is an additional reduction assumption:
higher connected functions are set to zero at coincident points in the
thermodynamic limit,
\[
C_{m_1\dots m_q}(n,n,\ldots)=0\qquad
  \text{whenever two indices coincide, for all } q>2,\; L\to\infty.
\]
This assumption is compatible with the scalar motivation discussed in
Ref.~\cite{Frasca:2026dyg} and with the role of triviality results in scalar
theories for $d\ge 4$~\cite{Aizenman:2019yuo}, but it is not a theorem about
non-Abelian Yang--Mills theory. It should therefore be regarded as a
controlled ansatz for exploring a reduced sector of the exact lattice
hierarchy. With this understanding, the continuum analogue of the closure
condition is written as
\[
G_n(x,x,\dots)=0,\qquad n>2.
\]
As a result, one obtains
\begin{eqnarray}
\lefteqn{i\eta_\mu^a\partial^2\phi(m)\ =\ igf_{abc}\Big[\eta^\nu_b\eta_\nu^c
  (\partial_\mu \phi(m))\phi(m)+\strut}\nonumber\\&&\strut\qquad
  -2\eta_\mu^b\eta_\nu^c(\partial^\nu \phi(m))\phi(m)
  +\eta_\nu^b\eta_\mu^c(\partial^\nu \phi(m))\phi(m)\Big]
  +\strut\nonumber\\&&\strut
  +ig^2f_{bcd}f_{dea}\Big[\eta_\mu^\nu\delta_e^b\eta_\nu^cG(m,m)\phi(m)
  +\eta_\nu^\nu\delta_e^c\eta_\mu^bG(m,m)\phi(m)
  +\eta_\mu^b\eta_\nu^c\eta^\nu_e\phi^3(m)\Big].
\end{eqnarray}
Contracting with $\eta^\mu_a$ and using the orthogonality relation
\begin{equation}\label{ortho}
\eta_\mu^a\eta^\mu_b=\delta^a_b,
\end{equation}
together with $f_{acd}f_{bcd}=N\delta_{ab}$ and
$f_{bcd}f_{dbc}=-f_{bcd}f_{dcb}=N(N^2-1)$ one obtains
\begin{eqnarray}
\lefteqn{i(N^2-1)\partial^2\phi(m)\ =\ igf_{abc}\Big[\eta^\mu_a\eta^\nu_b
  \eta_\nu^c(\partial_\mu\phi(m))\phi(m)+\strut}\nonumber\\&&\strut\qquad
  -2\eta^\mu_a\eta_\mu^b\eta_\nu^c(\partial^\nu\phi(m))\phi(m)
  +\eta^\mu_a\eta_\nu^b\eta_\mu^c(\partial^\nu\phi(m))\phi(m)\Big]
  +\strut\nonumber\\&&\strut
  +ig^2f_{bcd}f_{dea}\Big[\eta^\mu_a\eta_\mu^\nu\delta_e^b\eta_\nu^cG(m,m)
  \phi(m)+\eta^\mu_a\eta_\nu^\nu\delta_e^c\eta_\mu^bG(m,m)\phi(m)
  +\eta^\mu_a\eta_\mu^b\eta_\nu^c\eta^\nu_e\phi^3(m)\Big]
  \ =\kern-3pt\nonumber\\
  &=&igf_{abc}\Big[\eta^\mu_a\delta_b^c(\partial_\mu\phi(m))\phi(m)
  -2\delta_a^b\eta_\nu^c(\partial^\nu\phi(m))\phi(m)
  +\delta_a^c\eta_\nu^b(\partial^\nu\phi(m))\phi(m)\Big]
  +\strut\nonumber\\&&\strut
  +ig^2f_{bcd}f_{dea}\Big[\eta^\nu_a\delta_e^b\eta_\nu^cG(m,m)\phi(m)
  +D\eta^\mu_a\delta_e^c\eta_\mu^bG(m,m)\phi(m)
  +\delta_a^b\eta_\nu^c\eta^\nu_e\phi^3(m)\Big]\ =\nonumber\\
  &=&-g^2f_{bcd}f_{dea}\Big[\delta_a^c\delta_e^bG(m,m)\phi(m)
  +D\delta_a^b\delta_e^cG(m,m)\phi(m)+\delta_a^b\delta^c_e\phi^3(m)\Big]
  \ =\nonumber\\
  &=&ig^2\Big[f_{bcd}f_{dbc}G(m,m)\phi(m)+Df_{bcd}f_{dcb}G(m,m)\phi(m)
  +f_{bcd}f_{dcb}\phi^3(m)\Big]\ =\nonumber\\
  &=&iN(N^2-1)g^2\Big[(D-1)G(m,m)\phi(m)+\phi^3(m)\Big],
\end{eqnarray}
or
\begin{equation}\label{latt1}
\partial^2\phi(m)-Ng^2\Big[(D-1)G(m,m)\phi(m)+\phi^3(m)\Big]=0.
\end{equation}
Here $D=4$ is the space-time dimension; it is kept arbitrary in the algebra to
display the dimensional dependence of the coefficients. The same procedure
applied to the equation of motion for the two-point function leads to
\begin{eqnarray}
\lefteqn{4iD\delta_{mn}\ =\ -D\partial^2_mG(m,n)-(\partial^2\phi(m))\phi(n)
  \Big]+\strut}\nonumber\\&&\strut+Ng^2\Big[
  D(D-1)G(m,m)G(m,n)+(D-1)\phi^2(m)G(m,n)+\strut\nonumber\\&&\strut\qquad
  +(D-1)\phi(m)G(m,m)\phi(n)+\phi^3(m)\phi(n)\Big].
\end{eqnarray}
Using the previous result~(\ref{latt1}), this simplifies to
\begin{equation}\label{latt2}
4iD\delta_{mn}=-D\partial^2_mG(m,n)
  +Ng^2(D-1)\Big[DG(m,m)+\phi^2(m)\Big]G(m,n).
\end{equation}

\section{Discussion and Conclusion}
The distinction between the exact lattice result and the gauge-fixed
comparison is essential. The Haar-measure identity is an exact statement of
the unfixed compact-link theory, whereas the continuum and scalar equations
displayed here are ghost-omitted gluonic projections. A complete
covariant-gauge extension must restore the ghost sector and the associated
Slavnov-Taylor constraints.

We have shown that the Dyson--Schwinger hierarchy of Wilson lattice
Yang--Mills theory can be derived directly from the geometry of the compact
link variables. The essential input is the left invariance of the normalized
Haar measure. This produces an exact finite-lattice integration-by-parts
identity on the product of $\SU(N)$ group manifolds, with no appeal to a
formally flat gauge-potential measure and no need to introduce gauge fixing
merely to define the underlying lattice path integral.

The central technical outcome is the Wilson-action master equation written
link by link in terms of left-invariant Lie derivatives and staples. This
result is useful because it identifies the lattice origin of the functional
identities that are usually written in continuum language. In this form, the
hierarchy is anchored in the exact compact-group formulation of the theory,
while still being close enough to continuum notation to permit a controlled
comparison with gauge-fixed Dyson-Schwinger equations.

The continuum and scalar-reduced equations obtained in the later sections
should be read as conditional consequences of the exact lattice identity. The
continuum expansion introduces gauge-fixed gauge-field variables, and the
scalar closure imposes a special color-Lorentz ansatz together with a
coincident-point closure assumption. These steps are not part of the exact
theorem. Rather, they illustrate how the exact lattice master equation can be
projected into analytically tractable lower sectors, and they expose precisely
where additional dynamical assumptions enter.

The conceptual implication is that one can use the compact lattice theory not
merely as a numerical regulator, but also as a source of exact functional
identities from which continuum truncations may be engineered. This opens a
route toward truncation schemes that are less detached from the underlying
group-valued degrees of freedom. Such schemes could be valuable for studying
infrared Yang--Mills dynamics, mass generation, and the relation between
gauge-fixed Green functions and gauge-invariant observables.

Several concrete directions follow from this work. A first priority is to
complete the gauge-fixed continuum reduction by including the Faddeev--Popov
determinant, ghost correlators, and Slavnov--Taylor identities explicitly. A
second direction is to formulate a fully gauge-invariant Wilson-loop version
of the link-local master identity and compare it directly with loop equations.
A third direction is to analyze the renormalization of coincident quantities
such as $G(x,x)$ when the reduced equations are taken toward the continuum
limit. More broadly, the exact Haar-measure identity derived here provides a
stable starting point for lattice-informed functional methods in
nonperturbative gauge theory.

In this sense, our analysis establishes more than a formal rewriting of
Dyson--Schwinger equations. It provides a compact-group foundation for
connecting exact lattice identities, continuum functional methods, and
possible new truncation schemes. The finite-lattice identity is exact, the
continuum reduction is improvable, and the scalar closure is replaceable. This
modular structure is the main strength of the approach and the reason it may
be useful beyond the illustrative reductions studied here.

\section{Acknowledgments}
A.~G. acknowledges support from the Royal Society, UK,
funding reference: NIF\ R1\ 253963.

\begin{appendix}

\section{Lie Derivative of the Plaquette}
\setcounter{equation}{0}\def\theequation{A\arabic{equation}}
This appendix gives the explicit plaquette derivative used in Sec.~IV. The
gauge field $\xi_\mu^a$ is attached to the midpoint of the corresponding link,
and all shifted arguments label the midpoints of the links entering the two
plaquettes that contain the varied link $U_\mu(m)$. The expression is
displayed through cubic order in the small-field expansion, which is the order
needed to obtain the gluonic terms kept in the continuum reduction.

For readability we keep the formula in a fully expanded form. The three trace
structures below organize the result by order: the terms proportional to
$\Tr(T^aT^b)$ are linear in the field, those proportional to $\Tr(T^aT^bT^c)$
are quadratic, and those proportional to $\Tr(T^aT^bT^cT^d)$ are cubic. Terms
of higher order in $g$ are not displayed.

The Lie derivative of the plaquette for fixed directions $\hat\mu$ and
$\hat\nu$ on the lattice is given by
\begin{eqnarray}
\lefteqn{{\cal L}^a_{\mu,m} U_{\mu\nu}(m)
  \ :=\ i\Tr[T^aU_\mu(m)\Sigma_\mu^\dagger(m)
  -\Sigma_\mu(m)T^aU_\mu^\dagger(m)]\ =}\nonumber\\
  &=&2ga\Bigg\{\Bigg[
  \xi_\mu^b(m+\frac{\hat\mu}2+\hat\nu)-2\xi_\mu^b(m+\frac{\hat\mu}2)
  +\xi_\mu^b(m+\frac{\hat\mu}2-\hat\nu)\Bigg]+\strut\nonumber\\&&\strut
  -\Bigg[\xi_\nu^b(m+\hat\mu+\frac{\hat\nu}2)-\xi_\nu^b(m+\frac{\hat\nu}2)
  -\xi_\nu^b(m+\hat\mu-\frac{\hat\nu}2)+\xi_\nu^b(m-\frac{\hat\nu}2)\Bigg]
  \Bigg\}\Tr[T^aT^b]+\strut\nonumber\\&&\strut\kern-12pt
  -ig^2a^2\Bigg\{\Bigg[\xi_\mu^a(m+\frac{\hat\mu}2+\hat\nu)
  \left(\xi_\nu^c(m+\hat\mu+\frac{\hat\nu}2)+\xi_\nu^c(m+\frac{\hat\nu}2)
  \right)+\strut\nonumber\\&&\strut\qquad
  -\xi_\mu^b(m+\frac{\hat\mu}2-\hat\nu)\left(\xi_\nu^c(m+\hat\mu
  -\frac{\hat\nu}2)+\xi_\nu^c(m-\frac{\hat\nu}2)\right)\Bigg]
  +\strut\nonumber\\&&\strut
  -\Bigg[\left(\xi_\nu^b(m+\hat\mu+\frac{\hat\nu}2)+\xi_\nu^b(m+\frac{\hat\nu}2
  )\right)\xi_\mu^c(m+\frac{\hat\mu}2+\hat\nu)+\strut\nonumber\\&&\strut\qquad
  -\left(\xi_\nu^b(m+\hat\mu-\frac{\hat\nu}2)+\xi_\nu^b(m-\frac{\hat\nu}2)
  \right)\xi_\mu^c(m+\frac{\hat\mu}2-\hat\nu)\Bigg]+\strut\nonumber\\&&\strut
  -\Bigg[\xi_\nu^b(m+\hat\mu+\frac{\hat\nu}2)\xi_\nu^c(m+\frac{\hat\nu}2)
  -\xi_\nu^b(m+\frac{\hat\nu}2)\xi_\nu^c(m+\hat\mu+\frac{\hat\nu}2)
  +\strut\nonumber\\&&\strut\qquad
  +\xi_\nu^b(m+\hat\mu-\frac{\hat\nu}2)\xi_\nu^c(m-\frac{\hat\nu}2)
  -\xi_\nu^b(m-\frac{\hat\nu}2)\xi_\nu^c(m+\hat\mu-\frac{\hat\nu}2)\Bigg]
  \Bigg\}\Tr(T^aT^bT^c)+\strut\nonumber\\&&\strut\kern-12pt
  -\frac{g^3a^3}6\Bigg\{2\Bigg[
  \xi_\mu^b(m+\frac{\hat\mu}2+\hat\nu)\xi_\mu^c(m+\frac{\hat\mu}2+\hat\nu)
  \xi_\mu^d(m+\frac{\hat\mu}2+\hat\nu)+\strut\nonumber\\&&\strut\qquad
  -3\xi_\mu^b(m+\frac{\hat\mu}2)\xi_\mu^c(m+\frac{\hat\mu}2+\hat\nu)
  \xi_\mu^d(m+\frac{\hat\mu}2+\hat\nu)+\strut\nonumber\\&&\strut\qquad
  +3\xi_\mu^b(m+\frac{\hat\mu}2)\xi_\mu^c(m+\frac{\hat\mu}2)
  \xi_\mu^d(m+\frac{\hat\mu}2+\hat\nu)+\strut\nonumber\\&&\strut\qquad
  -2\xi_\mu^b(m+\frac{\hat\mu}2)\xi_\mu^c(m+\frac{\hat\mu}2)
  \xi_\mu^d(m+\frac{\hat\mu}2)+\strut\nonumber\\&&\strut\qquad
  +3\xi_\mu^b(m+\frac{\hat\mu}2)\xi_\mu^c(m+\frac{\hat\mu}2)
  \xi_\mu^d(m+\frac{\hat\mu}2-\hat\nu)+\strut\nonumber\\&&\strut\qquad
  -3\xi_\mu^b(m+\frac{\hat\mu}2)\xi_\mu^c(m+\frac{\hat\mu}2-\hat\nu)
  \xi_\mu^d(m+\frac{\hat\mu}2-\hat\nu)+\strut\nonumber\\&&\strut\qquad
  +\xi_\mu^b(m+\frac{\hat\mu}2-\hat\nu)\xi_\mu^c(m+\frac{\hat\mu}2-\hat\nu)
  \xi_\mu^d(m+\frac{\hat\mu}2-\hat\nu)\Bigg]+\strut\nonumber\\&&\strut
  -3\Bigg[\left(\xi_\nu^b(m+\hat\mu+\frac{\hat\nu}2)
  -\xi_\nu^b(m+\frac{\hat\nu}2)\right)\xi_\mu^c(m+\frac{\hat\mu}2+\hat\nu)
  \xi_\mu^d(m+\frac{\hat\mu}2+\hat\nu)+\strut\nonumber\\&&\strut\qquad
  -\left(\xi_\nu^b(m+\hat\mu-\frac{\hat\nu}2)-\xi_\nu^b(m-\frac{\hat\nu}2)
  \right)\xi_\mu^c(m+\frac{\hat\mu}2-\hat\nu)
  \xi_\mu^d(m+\frac{\hat\mu}2-\hat\nu)\Bigg]+\strut\nonumber\\&&\strut
  +6\Bigg[\xi_\mu^b(m+\frac{\hat\mu}2)\xi_\mu^d(m+\frac{\hat\mu}2+\hat\nu)
  \left(\xi_\nu^c(m+\hat\mu+\frac{\hat\nu}2)-\xi_\nu^c(m+\frac{\hat\nu}2)
  \right)+\strut\nonumber\\&&\strut\qquad
  -\xi_\mu^b(m+\frac{\hat\mu}2)\xi_\mu^d(m+\frac{\hat\mu}2-\hat\nu)
  \left(\xi_\nu^c(m+\hat\mu-\frac{\hat\nu}2)-\xi_\nu^c(m-\frac{\hat\nu}2)
  \right)\Bigg]+\strut\nonumber\\&&\strut
  +3\Bigg[\left(\xi_\nu^b(m+\hat\mu+\frac{\hat\nu}2)
  \xi_\nu^c(m+\hat\mu+\frac{\hat\nu}2)+\xi_\nu^b(m+\frac{\hat\nu}2)
  \xi_\nu^c(m+\frac{\hat\nu}2)\right)\xi_\mu^d(m+\frac{\hat\mu}2+\hat\nu)
  +\strut\nonumber\\&&\strut\qquad
  +\left(\xi_\nu^b(m+\hat\mu-\frac{\hat\nu}2)
  \xi_\nu^c(m+\hat\mu-\frac{\hat\nu}2)+\xi_\nu^b(m-\frac{\hat\nu}2)
  \xi_\nu^c(m-\frac{\hat\nu}2)\right)\xi_\mu^d(m+\frac{\hat\mu}2-\hat\nu)
  \Bigg]+\strut\nonumber\\&&\strut+3\Bigg[
  2\xi_\mu^b(m+\frac{\hat\mu}2)\left(\xi_\mu^c(m+\frac{\hat\mu}2+\hat\nu)
  -\xi_\mu^c(m+\frac{\hat\mu}2)\right)
  \left(\xi_\nu^d(m+\hat\mu+\frac{\hat\nu}2)-\xi_\nu^d(m+\frac{\hat\nu}2)
  \right)+\strut\nonumber\\&&\strut\qquad
  -\xi_\mu^b(m+\frac{\hat\mu}2+\hat\nu)\xi_\mu^c(m+\frac{\hat\mu}2+\hat\nu)
  \left(\xi_\nu^d(m+\hat\mu+\frac{\hat\nu}2)-\xi_\nu^d(m+\frac{\hat\nu}2)
  \right)+\strut\nonumber\\&&\strut\qquad
  -2\xi_\mu^b(m+\frac{\hat\mu}2)\left(\xi_\mu^c(m+\frac{\hat\mu}2-\hat\nu)
  -\xi_\mu^c(m+\frac{\hat\mu}2)\right)
  \left(\xi_\nu^d(m+\hat\mu-\frac{\hat\nu}2)-\xi_\nu^d(m-\frac{\hat\nu}2)
  \right)+\kern-10pt\strut\nonumber\\&&\strut\qquad
  +\xi_\mu^b(m+\frac{\hat\mu}2-\hat\nu)\xi_\mu^c(m+\frac{\hat\mu}2-\hat\nu)
  \left(\xi_\nu^d(m+\hat\mu-\frac{\hat\nu}2)-\xi_\nu^d(m-\frac{\hat\nu}2)
  \right)\Bigg]+\strut\nonumber\\&&\strut
  -6\Bigg[\xi_\mu^c(m+\frac{\hat\mu}2+\hat\nu)
  \left(\xi_\nu^b(m+\hat\mu+\frac{\hat\nu}2)\xi_\nu^d(m+\frac{\hat\nu}2)
  +\xi_\nu^b(m+\frac{\hat\nu}2)\xi_\nu^d(m+\hat\mu+\frac{\hat\nu}2)\right)
  +\strut\nonumber\\&&\strut\qquad
  +\xi_\mu^c(m+\frac{\hat\mu}2-\hat\nu)
  \left(\xi_\nu^b(m+\hat\mu-\frac{\hat\nu}2)\xi_\nu^d(m-\frac{\hat\nu}2)
  +\xi_\nu^b(m-\frac{\hat\nu}2)\xi_\nu^d(m+\hat\mu-\frac{\hat\nu}2)\right)
  \Bigg]+\strut\nonumber\\&&\strut
  +3\Bigg[\left(\xi_\mu^b(m+\frac{\hat\mu}2+\hat\nu)
  -2\xi_\mu^b(m+\frac{\hat\mu}2)\right)
  \times\nonumber\\&&\strut\qquad\qquad\times
  \left(\xi_\nu^c(m+\hat\mu+\frac{\hat\nu}2)
  \xi_\nu^d(m+\hat\mu+\frac{\hat\nu}2)+\xi_\nu^c(m+\frac{\hat\nu}2)
  \xi_\nu^d(m+\frac{\hat\nu}2)\right)+\strut\nonumber\\&&\strut\qquad
  +2\xi_\mu^b(m+\frac{\hat\mu}2)\left(\xi_\nu^c(m+\hat\mu+\frac{\hat\nu}2)
  \xi_\nu^d(m+\frac{\hat\nu}2)+\xi_\nu^c(m+\frac{\hat\nu}2)
  \xi_\nu^d(m+\hat\mu+\frac{\hat\nu}2)\right)+\strut\nonumber\\&&\qquad
  +\left(\xi_\mu^b(m+\frac{\hat\mu}2-\hat\nu)
  -2\xi_\mu^b(m+\frac{\hat\mu}2)\right)
  \times\nonumber\\&&\strut\qquad\qquad\times
  \left(\xi_\nu^c(m+\hat\mu-\frac{\hat\nu}2)
  \xi_\nu^d(m+\hat\mu-\frac{\hat\nu}2)+\xi_\nu^c(m-\frac{\hat\nu}2)
  \xi_\nu^d(m-\frac{\hat\nu}2)\right)+\strut\nonumber\\&&\strut\qquad
  +2\xi_\mu^b(m+\frac{\hat\mu}2)\left(\xi_\nu^c(m+\hat\mu-\frac{\hat\nu}2)
  \xi_\nu^d(m-\frac{\hat\nu}2)+\xi_\nu^c(m-\frac{\hat\nu}2)
  \xi_\nu^d(m+\hat\mu-\frac{\hat\nu}2)\right)\Bigg]+\strut\nonumber\\&&\strut
  -\Bigg[2\xi_\nu^b(m+\hat\mu+\frac{\hat\nu}2)
  \xi_\nu^c(m+\hat\mu+\frac{\hat\nu}2)\xi_\nu^d(m+\hat\mu+\frac{\hat\nu}2)
  +\strut\nonumber\\&&\strut\qquad
  -3\xi_\nu^b(m+\frac{\hat\nu}2)
  \xi_\nu^c(m+\hat\mu+\frac{\hat\nu}2)\xi_\nu^d(m+\hat\mu+\frac{\hat\nu}2)
  +\strut\nonumber\\&&\strut\qquad
  +3\xi_\nu^b(m+\frac{\hat\nu}2)\xi_\nu^c(m+\frac{\hat\nu}2)
  \xi_\nu^d(m+\hat\mu+\frac{\hat\nu}2)+\strut\nonumber\\&&\strut\qquad
  -3\xi_\nu^b(m+\hat\mu+\frac{\hat\nu}2)
  \xi_\nu^c(m+\hat\mu+\frac{\hat\nu}2)\xi_\nu^d(m+\frac{\hat\nu}2)
  +\strut\nonumber\\&&\strut\qquad
  +3\xi_\nu^b(m+\hat\mu+\frac{\hat\nu}2)\xi_\nu^c(m+\frac{\hat\nu}2)
  \xi_\nu^d(m+\frac{\hat\nu}2)+\strut\nonumber\\&&\strut\qquad
  -2\xi_\nu^b(m+\frac{\hat\nu}2)\xi_\nu^c(m+\frac{\hat\nu}2)
  \xi_\nu^d(m+\frac{\hat\nu}2)+\strut\nonumber\\&&\strut\qquad
  +2\xi_\nu^b(m-\frac{\hat\nu}2)\xi_\nu^c(m-\frac{\hat\nu}2)
  \xi_\nu^d(m-\frac{\hat\nu}2)+\strut\nonumber\\&&\strut\qquad
  -3\xi_\nu^b(m+\hat\mu-\frac{\hat\nu}2)\xi_\nu^c(m-\frac{\hat\nu}2)
  \xi_\nu^d(m-\frac{\hat\nu}2)+\strut\nonumber\\&&\strut\qquad
  +3\xi_\nu^b(m+\hat\mu-\frac{\hat\nu}2)
  \xi_\nu^c(m+\hat\mu-\frac{\hat\nu}2)\xi_\nu^d(m-\frac{\hat\nu}2)
  +\strut\nonumber\\&&\strut\qquad
  -3\xi_\nu^b(m-\frac{\hat\nu}2)\xi_\nu^c(m-\frac{\hat\nu}2)
  \xi_\nu^d(m+\hat\mu-\frac{\hat\nu}2)+\strut\nonumber\\&&\strut\qquad
  +3\xi_\nu^b(m-\frac{\hat\nu}2)
  \xi_\nu^c(m+\hat\mu-\frac{\hat\nu}2)\xi_\nu^d(m+\hat\mu-\frac{\hat\nu}2)
  +\strut\nonumber\\&&\strut\qquad
  -2\xi_\nu^b(m+\hat\mu-\frac{\hat\nu}2)
  \xi_\nu^c(m+\hat\mu-\frac{\hat\nu}2)\xi_\nu^d(m+\hat\mu-\frac{\hat\nu}2)
  \Bigg]\Bigg\}\Tr[T^aT^bT^cT^d],
\end{eqnarray}
where summation over the color indices $a$, $b$, $c$, and $d$ is implied. The
localization limit used in the main text is obtained by identifying all
shifted midpoint fields with a common value. In that limit the linear and
quadratic finite-difference pieces collapse as expected, and the cubic part
reduces to the double-commutator/color-factor structure shown below
Eq.~(\ref{master}). This provides a compact check on the signs and ordering of
the cubic contribution.

\end{appendix}

\end{document}